\documentclass{qxLab}

\usepackage{cite}
\usepackage{amsmath,amssymb,amsfonts}
\usepackage{algorithmic}
\usepackage{graphicx}
\usepackage{textcomp}
\usepackage{xcolor}
\usepackage{url}
\usepackage{listings}

\begin{document}

\title{How deep is your encoder: an analysis of features descriptors for an autoencoder-based audio-visual quality metric\\
\thanks{This publication has emanated from research supported in part by the Conselho Nacional de Desenvolvimento Cientfico e Tecnol\'ogico (CNPq), the Coordena\c{c}\~{a}o de Aperfei\c{c}oamento de Pessoal de N\'ivel Superior (CAPES), the Funda\c c\~ao de Apoio \`a Pesquisa do Distrito Federal (FAPDF), the University of Bras\'ilia (UnB), the research grant from Science Foundation Ireland (SFI) and the European Regional Development Fund under Grant Number 12/RC/2289\_P2 and Grant Number 13/RC/2077.}
}

\author{
  Helard Martinez \\
  University College Dublin\\
  Dublin, Ireland \\
  \texttt{helard.becerra@ucd.ie} \\
   \And
 Andrew Hines \\
  University College Dublin\\
  Dublin, Ireland \\
  \texttt{andrew.hines@ucd.ie} \\
     \And
 Myl\`ene C.Q. Farias \\
  University of Brasilia\\
  Brasilia, Brazil \\
  \texttt{mylene@ieee.org} \\
}

\maketitle

\begin{abstract}
The development of audio-visual quality assessment models poses a number of challenges in order to obtain accurate predictions. One of these challenges is the modelling of the complex interaction that audio and visual stimuli have and how this interaction is interpreted by human users. The No-Reference Audio-Visual Quality Metric Based on a Deep Autoencoder (NAViDAd) deals with this problem from a machine learning perspective. The metric receives two sets of audio and video features descriptors and produces a low-dimensional set of features used to predict the audio-visual quality. A basic implementation of NAViDAd was able to produce accurate predictions {tested with a range of} different audio-visual databases. The current work performs an ablation study on the base architecture of the metric. Several modules are removed or re-trained using different configurations to have a better understanding of the metric functionality. The results presented in this study provided important feedback that allows us to understand the real capacity of the metric’s architecture and eventually develop a much better audio-visual quality metric.
\end{abstract}

\keywords{audiovisual \and quality metrics \and autoencoder \and qoe \and machine learning}

\section{Introduction}\label{sec:intro}

Multimedia services and applications, such as digital television, IP-based video transmission, video gaming, social networks, and mobile services, have gained enormous relevance in the past years. Due to the advances in information and communication technologies and the great demand for audio-visual content, it is expected that the number of available services increase in the following years. Therefore, guaranteeing certain levels of quality of experience (QoE) to the end-user is key for the success of any multimedia service or application \cite{Korhonen2010}. In this context, the development of efficient tools, that are able to monitor and quantify the audio-visual experience (as perceived by the end-user), will provide with reliable information that permit reallocation of network resources in real-time. This will certainly bring benefits to Internet Service Providers (ISP), broadcast companies, application developers and common users \cite{akhtar2019multimedia}.

Over the last decades, researchers have developed different objective metrics (computational algorithms) capable of predicting the perceived quality of signal stimuli \cite{akhtar2017audio}. The performance of these metrics is estimated by comparing their predictions with the scores gathered at psychophysical experiments (subjective scores). {Commonly, objective metrics use relevant information from both original and transmitted signals and compare them in order to predict the perceived quality (Full and Reduced Reference metrics). For real-time monitoring purposes, metrics that require no information from the original signal (No-Reference metrics) are highly desired and an active research topic \cite{chikkerur2011objective}.} At present, the vast majority of proposed objective metrics deals with the quality assessment problem at a single-modal level, i.e., audio and video quality measured separately \cite{akhtar2017audio}. Several metrics have achieved high levels of performance by using sets of descriptive features \cite{liu2011multi,Malfait2006a}. A key factor in the success of these metrics is the ability of the extracted features to describe the audio and visual signals. In contrast, the multi-modal quality assessment problem, more specifically for audio-visual content, faces several challenges {in a No-Reference (NR) context}. One important issue is how to represent, (in terms of features) the complex interaction between audio and visual sensory channels~\cite{pinson2011audiovisual}. Since current proposals are limited to combining individual audio and video quality estimations and/or using a parametric approach, which depends entirely on the network parameters, the quality assessment of audio-visual signals remains as a relevant research topic \cite{martinez2018combining,Garcia2011a}.

\begin{figure}[t!]
\centering
\includegraphics[scale=0.45]{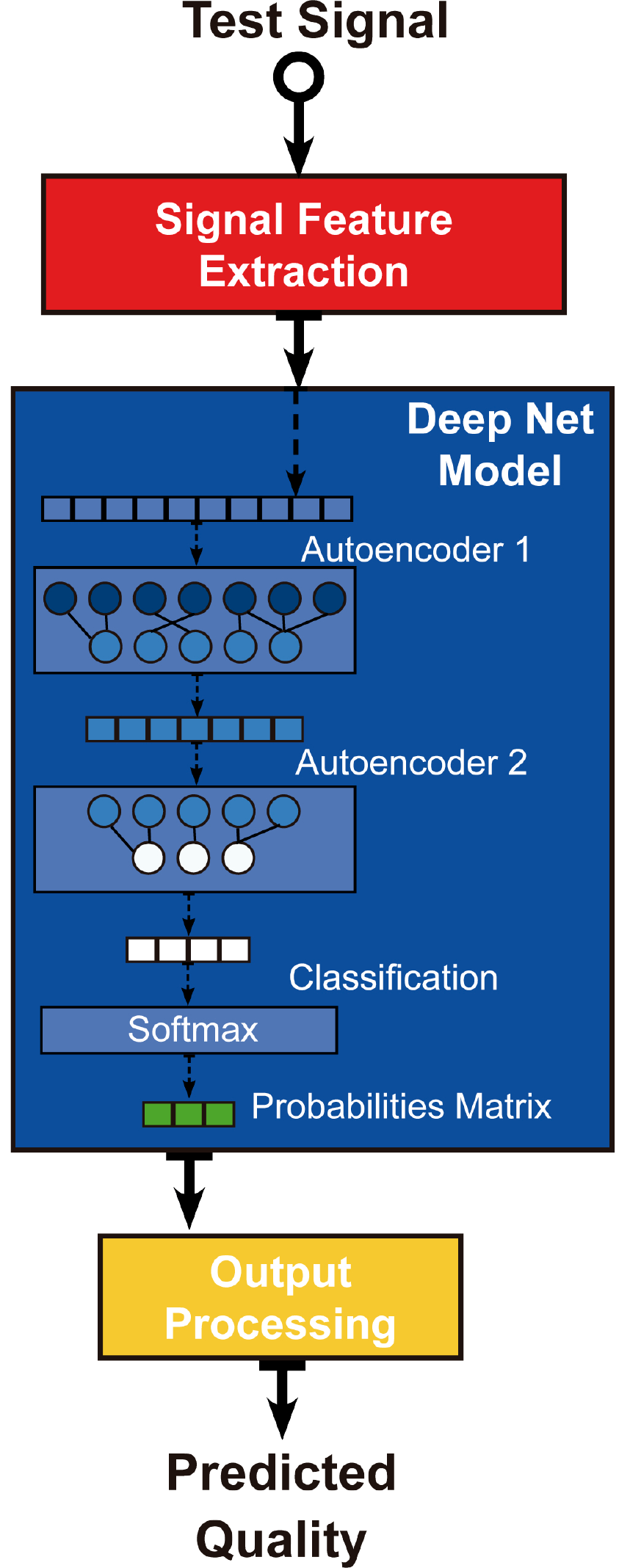}
\caption{Block diagram of the Audio-Visual Quality metric NAViDAd.}\label{fig:block_diagram}
\end{figure}

Machine learning (ML) has provided the quality assessment area with new tools and techniques to develop more accurate models. Several audio and video quality metrics {rely on} ML techniques to imitate different aspects of human perception, instead of modelling complex non-linear functions \cite{redi2010color}. {By exploiting this new set of capabilities, researchers have been able to develop several NR quality metrics, some of them to predict the quality of image and video content \cite{zhang2014c,mittal2012no,mittal2016completely}. These metrics are usually designed to extract descriptive features from the transmitted signal and then map these features into quality scores using a ML technique.} Other metrics have obtained stronger descriptive features by exploiting the capacity of ML techniques to capture the fundamental properties of audio and video signals \cite{soni2016novel}. {In line with this, an architecture that integrates these techniques to blindly estimate the perceived quality of a transmitted signal resulted in three NR metrics for audio, video and audio-visual content \cite{phdthesisHelard2019,martinez2019no,martinez2019navidad}. From this set of metrics,} the No-reference Audio-Visual Quality metric based on a Deep Autoencoder (NAViDAd) \cite{martinez2019navidad} {is one of the first attempts to develop a NR quality metric for audio-visual content.} The metric takes as input sets of audio and video features, extracted from the audio-visual signal under test. These sets of features are passed to a trained deep neural network, which is composed of two main modules: an autoencoder  and a classification. The autoencoder module takes the audio and video descriptive features and produces a new, low-dimensional, set of audio-visual descriptive features. NAViDAd exploits the capability of autoencoders to find strong relationships among input features and produce a new set of features that can describe the complex interactions between audio and video stimulus. Then, the classification module maps the new features into audio-visual quality scores. Finally, the metric scales the quality scores into a $<$1,5$>$ range. A block diagram of the NAViDAd metric is presented in Figure \ref{fig:block_diagram}.


A basic implementation of NAViDAd performed well when tested on different audio-visual databases. However, a more detailed analysis is required to fully understand the real impact of certain modules (and their configuration) in the audio-visual quality prediction. In the context of ML, ablation studies represent a straightforward method to understand {the relative importance of the different components (and their configurations) of a deep network \cite{meyes2019ablation}.}  By executing ablation experiments researchers are able to observe how the network performance is affected when certain modules are removed (or replaced). In this work, we present several ablation experiments executed in the baseline architecture of the NAViDAd metric. New models were re-trained using different configurations to have a better understanding of the metric functionality. The ablation experiments were organized in four main categories:
\begin{itemize}
    \item Internal layers, where layers were added or removed from the basic autoencoder structure {to verify how much effort during training (more or fewer layers) is accepted to get good performance.}
    \item Neuron’s distribution, where the dimension of the low-dimensional set of features was altered {to determine the level of compression (output dimension) that the autoencoder needed to provide to produce strong features.}
    \item Features selection, where either just video or audio features were used to train the model {to see how good they were at (individually) describing the combination of audio and video distortions.}
    \item Mapping strategy, where the classification module was replaced by a regression module {to evaluate if a continuous mapping (regression) was a better approach}.
\end{itemize}


The main objective of this work is to obtain important feedback that allows us to understand the real capacity of the metric’s architecture and eventually develop a much better audio-visual quality metric.

The structure of this paper is organized as follows. In Section \ref{sec:architecture}, the baseline training details of the NAViDAd metric is presented. In Section \ref{sec:ablation}, details from the ablation experiments are presented. In Section \ref{sec:discussion}, the ablation results are presented and discussed. Finally, in Section \ref{sec:conclusions} conclusions and final comments are presented.

\section{NAViDAd Baseline Training}\label{sec:architecture}

This section presents details about the training of the baseline version of the NAViDAd metric, such as the audio-visual material used for training, descriptive audio and video features and basic configuration of the model. 
NAViDAd  was trained using audio-visual material from the UnB-AVQ database\footnote{This database is available for download from the site of the University of Bras\'ilia (\url{www.ene.unb.br/mylene/databases.html}) and at The Consumer Digital Video Library ({\url{www.cdvl.org}} - create an account and search for UNB).} \cite{becerra2020Unb}. 
This  database  contains audio-visual sequences with their corresponding subjective scores. The database is composed of 800 audio-visual sequences with visual and audio distortions combined. Visual distortions include Bitrate Compression, Packet-Loss errors and Frame-Freezing. Meanwhile, the audio distortions are Background Noise, Chop effects, Clipping and Echo effects.

Prior training, sets of visual and audio descriptive features are extracted from the audio-visual sequences. The visual features are composed of 88 Natural Scene Statistics (NSS) features, obtained using the feature extract function from the DIIVINE metric \cite{zhang2014c} and the spatial and temporal indices~\cite{ostaszewska2007quantifying}. These features are extracted for each frame in the video and they form a 90-by-$m$ matrix (m is the number of video frames), which represents the set of descriptive visual features. The set of audio features is constructed using an intensity spectrogram representation of the audio signal, which is the same one used by the VISQOLAudio audio quality metric \cite{sloan2017objective}. Using the extracting function from VISQOLAudio, a 25-by-$m$ matrix is built, representing the set of  descriptive  audio features. Finally, both sets are properly scaled and merged together to form a 115-by-$m$ matrix, which is the training input of the model. Additionally, a target set is constructed using the subjective scores of each audio-visual signal. This target set serves as input during the training of the mapping function in the classification layer of the metric.

\begin{figure}[tb!]
\centering
\includegraphics[scale=0.4]{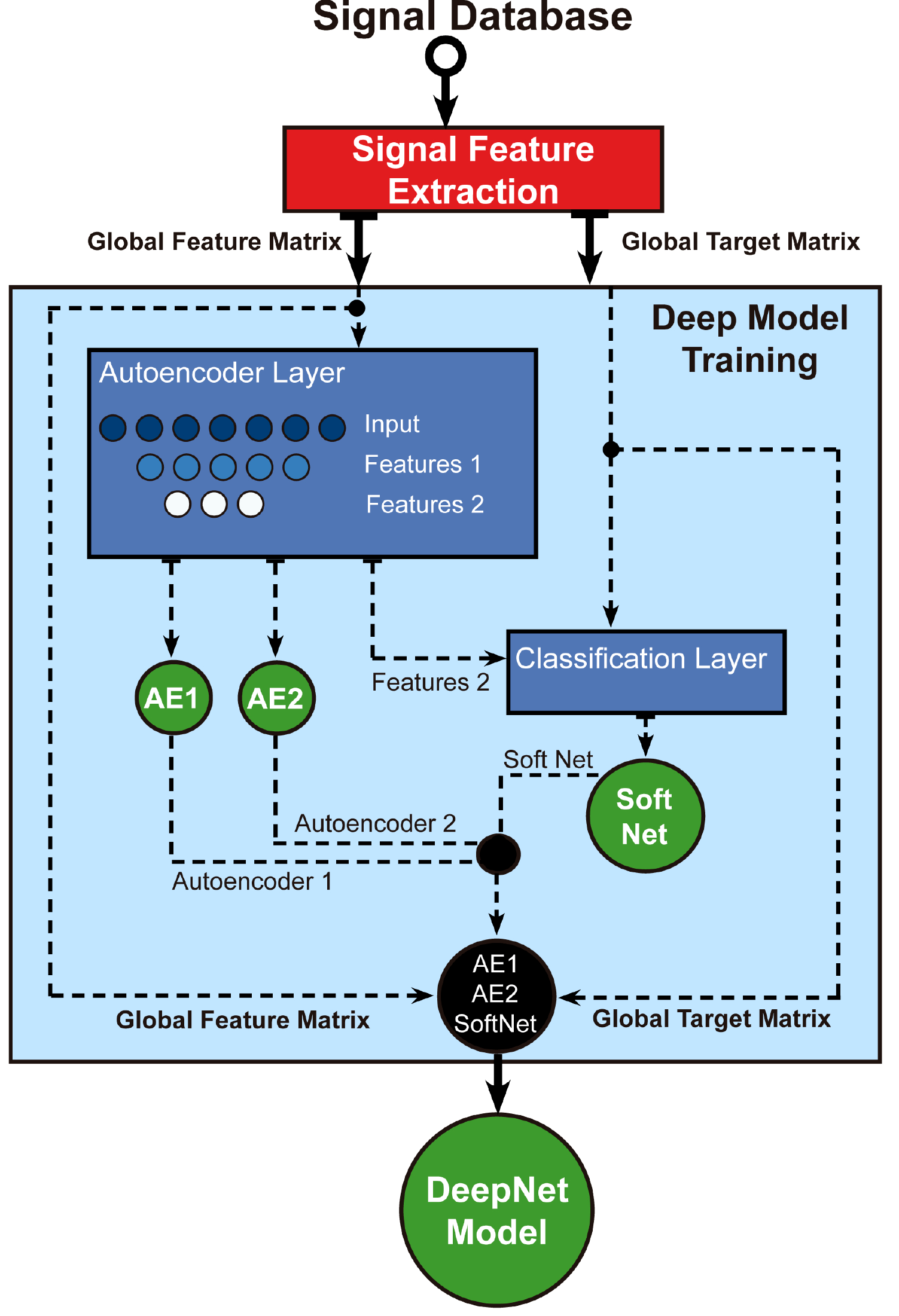}
\caption{Block diagram of the training stages of the NAViDAd metric.\label{fig:diagramAEModel}}
\end{figure}
\begin{table}[tb]
  \centering
  \caption{Parameters used for training the baseline version of NAViDAd.}
  \resizebox{0.5\columnwidth}{!}{  
    \begin{tabular}{rlr}
    \hline
    \multicolumn{1}{l}{\textbf{Module}} & \textbf{Parameters} & \textbf{Baseline - NAViDAd} \\
    \hline
    \multicolumn{1}{l}{\textbf{Autoencoder Module}} & \textbf{Input} & 115-by-$m$ matrix \\
          & \textbf{Internal layer size \#1} & 60 \\
          & \textbf{Internal layer size \#2} & 25 \\
          & \textbf{Decoder transfer function} & Linear \\
          & \textbf{L2 weigth regularization} & 0.001 \\
          & \textbf{Sparsity Regularization} & 4 \\
          & \textbf{Sparsity Proportion} & 0.05 \\
    \multicolumn{1}{l}{\textbf{Classification Module}} & \textbf{Input} & 25-by-$m$ matrix \\
          &       & 4-by-$m$ matrix \\
          & \textbf{Loss Function} & Cross Entropy \\
    \hline
    \multicolumn{1}{l}{\textbf{Additional Info}} & \textbf{Training Set} & UnB-AVQ (Experiment 3) \\
          & \textbf{\# sequences} & 800 \\
          & \textbf{Method} & 10-fold CV \\
    \hline
    \end{tabular}}%
  \label{tab:params}%
\end{table}%

Figure \ref{fig:diagramAEModel} shows the block diagram of NAViDAd baseline training. The baseline training is composed of two main modules: an autoencoder module and a classification module. The autoencoder module is trained using as input the extracted audio and video sets of features (115-by-$m$ matrix). This module transforms its input into a low-dimensional set of features with the capacity of describing the audio-visual characteristics of the signal. The basic configuration of the autoencoder module uses two internal layers with a neuron’s distribution of 115-60-25. The first layer (denoted as Autoencoder 1) receives as input the 115-by-$m$ matrix and  produces an output in the form of a 60-by-$m$ matrix (denoted as Features 1). Then, this output matrix is  input to train the second layer (denoted as Autoencoder 2). The second layer  produces a 25-by-$m$ matrix (denoted as Features 2), which represents the low-dimensional set of audio-visual features of the  signal. 

The classification module is trained using the low-dimensional set of audio-visual features (25-by-$m$ matrix) and the target set containing the subjective scores of the audio-visual signal. The classification module maps the low-dimensional set of features into its corresponding subjective score. To this end, a softmax function is used to identify the quality group that corresponds to the set of input features, the result is a trained function denoted as SoftNet. Finally, all trained modules: Autoencoder1, Autoencoder 2 and SoftNet are stacked to form the deep network. Table \ref{tab:params} presents additional information regarding the parameters used for training this basic model.

\begin{figure*}[h!]
\centering
\begin{tabular}{ccc}
\includegraphics[width = 0.32\columnwidth]{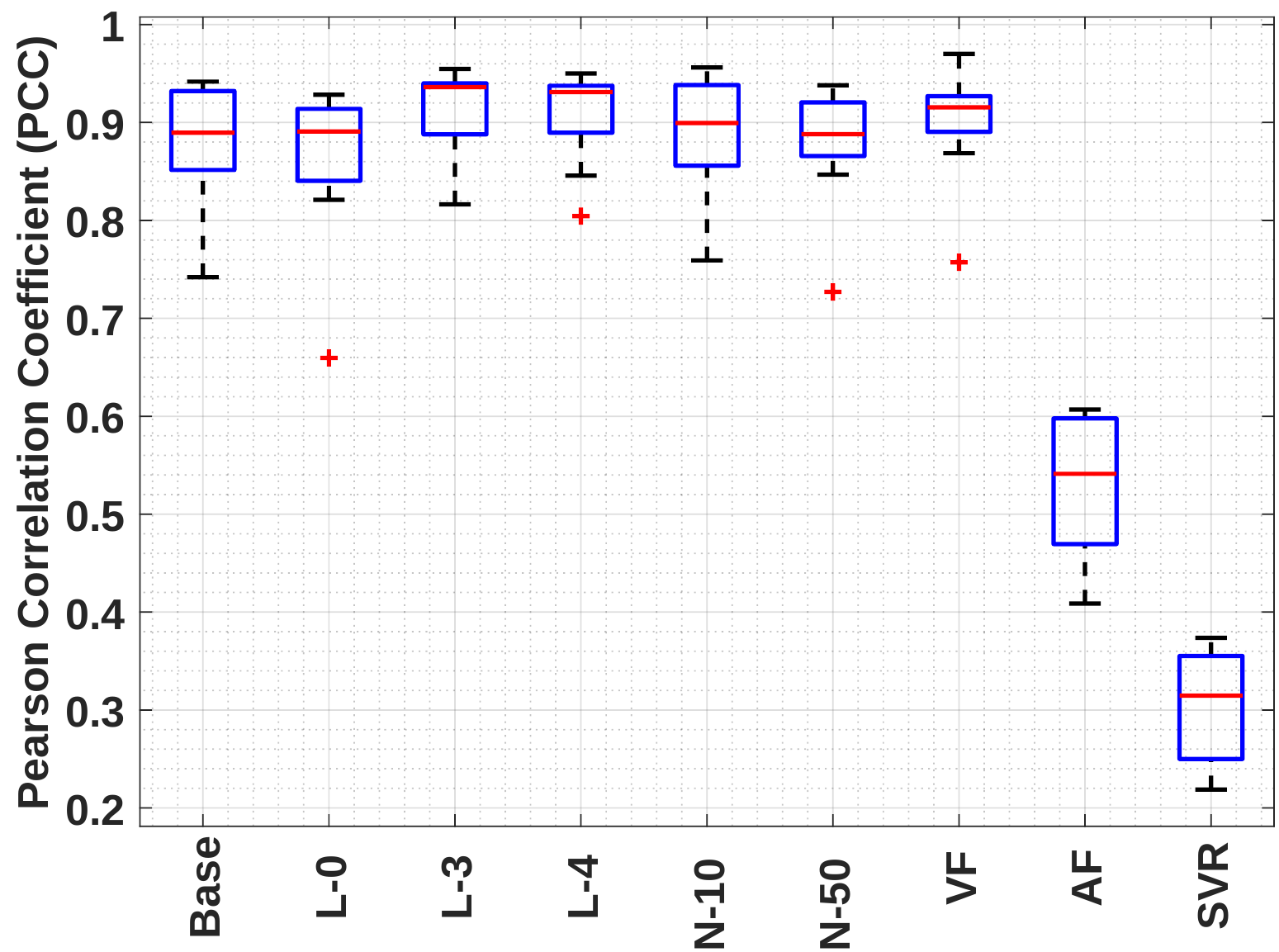} & 
\includegraphics[width = 0.32\columnwidth]{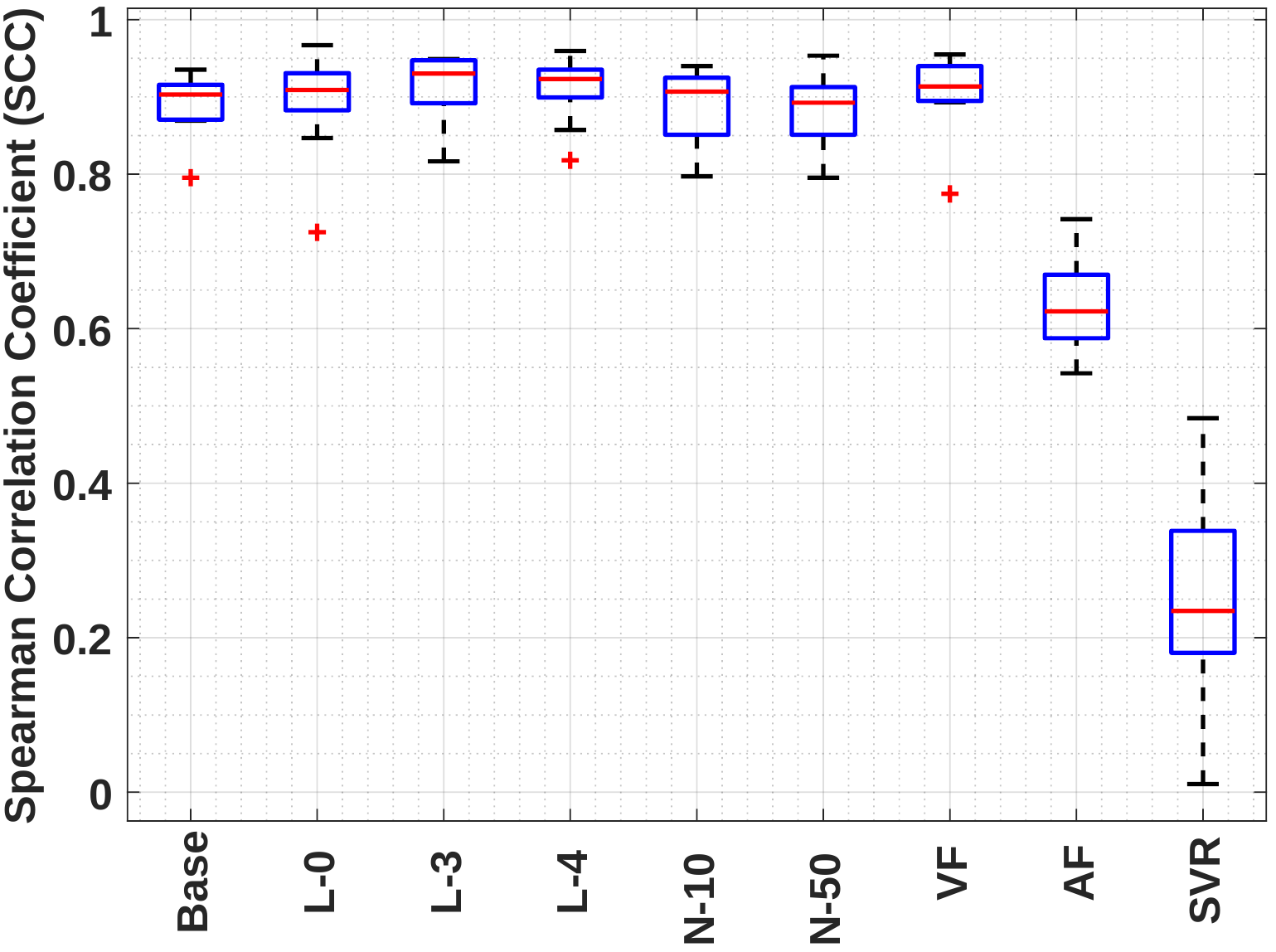} &
\includegraphics[width = 0.32\columnwidth]{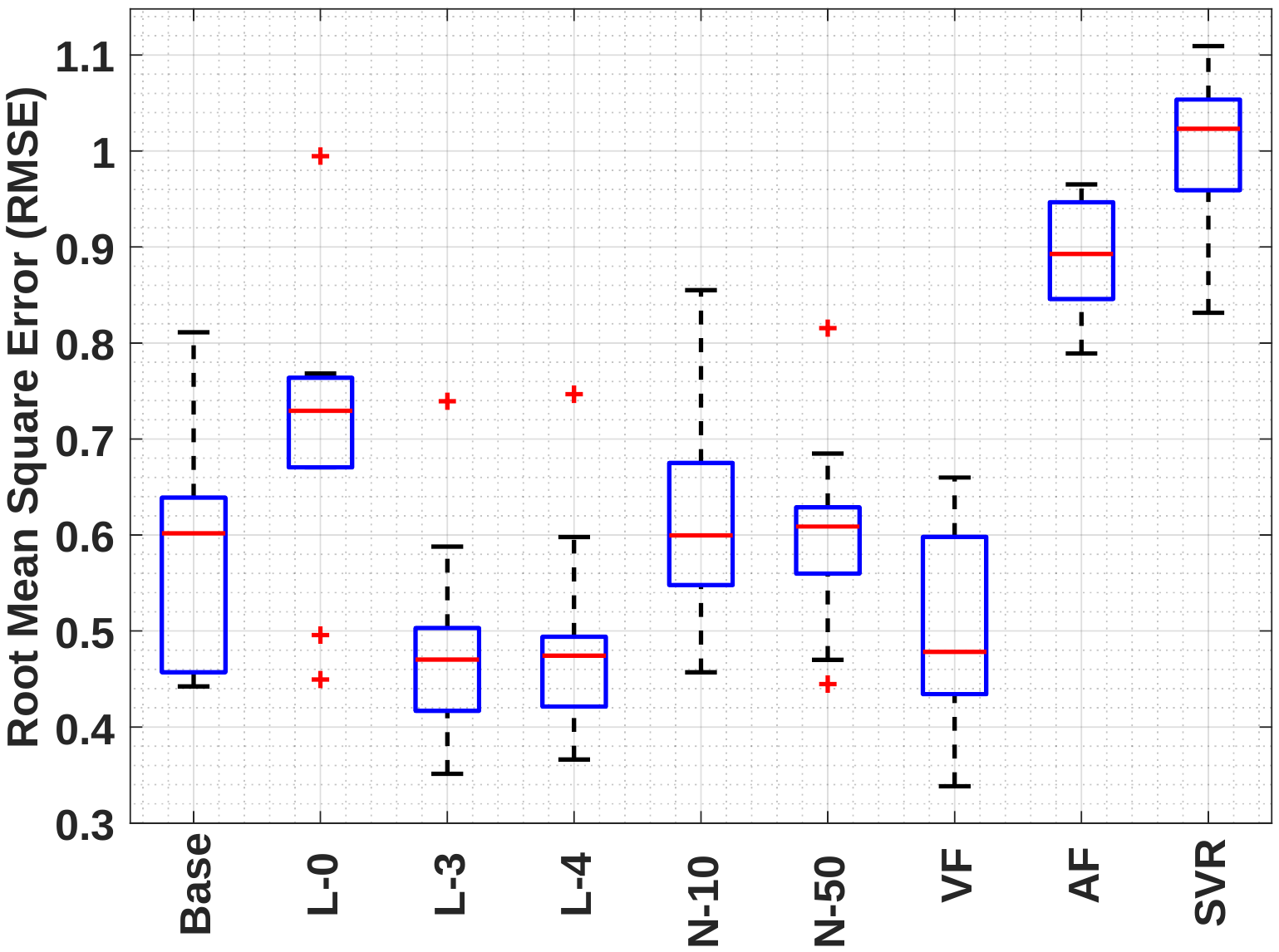} \\
\end{tabular}
\caption{PCC, SCC, and RMSE values of the predicted quality scores, for the different experiments.}\label{fig:ablation}
\end{figure*}
\begin{table*}[bt!]
  \centering
  \caption{Pearson and Spearman Correlation Coefficients (PCC and SCC) and Root Mean Square Error (RMSE) gathered from ablation experiments on UnB-AVQ database \cite{becerra2020Unb}.}
    \resizebox{0.9\columnwidth}{!}{
    \begin{tabular}{rl|rrrr|rr|r}
    \hline
          &       & \multicolumn{4}{c|}{\textbf{Audio Degradations}} & \multicolumn{2}{c|}{\textbf{Video Degradations}} & \multicolumn{1}{c}{{\textbf{Overall}}} \\
\cline{1-8}    \multicolumn{1}{l}{\textbf{Model}} & \textbf{Measure} & \multicolumn{1}{l}{\textbf{Noise}} & \multicolumn{1}{l}{\textbf{Chop}} & \multicolumn{1}{l}{\textbf{Clip}} & \multicolumn{1}{l|}{\textbf{Echo}} & \multicolumn{1}{l}{\textbf{Packet-Loss}} & \multicolumn{1}{l|}{\textbf{Frame-Freezing}} &  \\
    \hline
    \multicolumn{1}{l}{Baseline (115-60-25)} & PCC   & 0.8879 & 0.9252 & 0.8794 & 0.9044 & 0.8638 & 0.9167 & \textbf{0.8819} \\
    \multicolumn{1}{l}{2AE, Softmax, AV Features} & SCC   & 0.9200 & 1.0000 & 0.8629 & 0.9086 & 0.8773 & 0.9050 & \textbf{0.8904} \\
          & RMSE  & 0.5764 & 0.6125 & 0.5406 & 0.6013 & 0.5931 & 0.5718 & \textbf{0.5850} \\
    \hline
    \multicolumn{1}{l}{\textbf{Internal Layers}} &       &       &       &       &       &       &       &  \\
    \hline
    \multicolumn{1}{l}{Layers-0 (115)} & PCC   & 0.8329 & 0.9110 & 0.8648 & 0.9091 & 0.8478 & 0.9011 & \textbf{0.8640} \\
    \multicolumn{1}{l}{0AE, Softmax, AV Features} & SCC   & 0.8500 & 1.0000 & 0.8971 & 0.8914 & 0.8782 & 0.8467 & \textbf{0.8926} \\
          & RMSE  & 0.7032 & 0.7544 & 0.6449 & 0.7196 & 0.6867 & 0.7212 & \textbf{0.7048} \\
    \hline
    \multicolumn{1}{l}{Layers-3 (115-90-40-25)} & PCC   & 0.9122 & 0.9640 & 0.9137 & 0.9262 & 0.8971 & 0.9381 & \textbf{0.9135} \\
    \multicolumn{1}{l}{3AE, Softmax, AV Features} & SCC   & 0.9200 & 1.0000 & 0.8857 & 0.9200 & 0.8973 & 0.9117 & \textbf{0.9137} \\
          & RMSE  & 0.5096 & 0.4368 & 0.4750 & 0.4828 & 0.5012 & 0.4688 & \textbf{0.4885} \\
    \hline
    \multicolumn{1}{l}{Layers-4 (115-100-70-50-25)} & PCC   & 0.9096 & 0.9583 & 0.9128 & 0.9248 & 0.8920 & 0.9372 & \textbf{0.9101} \\
    \multicolumn{1}{l}{4AE, Softmax, AV Features} & SCC   & 0.9100 & 1.0000 & 0.8743 & 0.9314 & 0.8927 & 0.9083 & \textbf{0.9096} \\
          & RMSE  & 0.5183 & 0.4332 & 0.4766 & 0.4782 & 0.5014 & 0.4734 & \textbf{0.4910} \\
    \hline
    \multicolumn{1}{l}{\textbf{Neurons Distribution}} &       &       &       &       &       &       &       &  \\
    \hline
    \multicolumn{1}{l}{Nodes-10 (115-60-10)} & PCC   & 0.8982 & 0.9177 & 0.8902 & 0.9203 & 0.8648 & 0.9216 & \textbf{0.8887} \\
    \multicolumn{1}{l}{2AE, Softmax, AV Features} & SCC   & 0.9200 & 1.0000 & 0.8686 & 0.9314 & 0.8782 & 0.9033 & \textbf{0.8881} \\
          & RMSE  & 0.6167 & 0.6556 & 0.5760 & 0.6176 & 0.6160 & 0.6153 & \textbf{0.6177} \\
    \hline
    \multicolumn{1}{l}{Nodes-50 (115-60-50)} & PCC   & 0.8828 & 0.9175 & 0.8754 & 0.9053 & 0.8555 & 0.9162 & \textbf{0.8778} \\
    \multicolumn{1}{l}{2AE, Softmax, AV Features} & SCC   & 0.9200 & 1.0000 & 0.8743 & 0.8857 & 0.8573 & 0.9000 & \textbf{0.8857} \\
          & RMSE  & 0.6040 & 0.6375 & 0.5549 & 0.6111 & 0.6050 & 0.5948 & \textbf{0.6026} \\
    \hline
    \multicolumn{1}{l}{\textbf{Features Selection}} &       &       &       &       &       &       &       &  \\
    \hline
    \multicolumn{1}{l}{VF (90-50-20)} & PCC   & 0.9108 & 0.9526 & 0.8962 & 0.9154 & 0.8806 & 0.9308 & \textbf{0.9032} \\
    \multicolumn{1}{l}{2AE, Softmax, V Features} & SCC   & 0.9300 & 1.0000 & 0.8857 & 0.9257 & 0.8682 & 0.8600 & \textbf{0.9063} \\
          & RMSE  & 0.4691 & 0.4604 & 0.5123 & 0.4905 & 0.5269 & 0.4580 & \textbf{0.4978} \\
    \hline
    \multicolumn{1}{l}{AF (25-18-10)} & PCC   & 0.7234 & 0.9827 & 0.9054 & 0.6581 & 0.4183 & 0.7567 & \textbf{0.5241} \\
    \multicolumn{1}{l}{2AE, Softmax, A Features} & SCC   & 0.6700 & 1.0000 & 0.8204 & 0.7537 & 0.6151 & 0.8251 & \textbf{0.6264} \\
          & RMSE  & 0.8059 & 1.1336 & 0.6733 & 0.9876 & 0.9452 & 0.8044 & \textbf{0.8895} \\
    \hline
    \multicolumn{1}{l}{\textbf{Mapping Strategy}} &       &       &       &       &       &       &       &  \\
    \hline
    \multicolumn{1}{l}{SVR (115-60-25)} & PCC   & 0.4989 & 0.8505 & 0.3064 & 0.3478 & 0.4845 & 0.9394 & \textbf{0.3056} \\
    \multicolumn{1}{l}{2AE, SVR, AV Features} & SCC   & 0.2800 & 0.8000 & 0.4457 & 0.2800 & 0.5482 & 0.8367 & \textbf{0.2496} \\
          & RMSE  & 0.9027 & 1.0836 & 0.9899 & 1.0246 & 1.0069 & 0.9864 & \textbf{0.9985} \\
    \hline
    \end{tabular}}%
  \label{tab:results}%
\end{table*}%

\section{Ablation Study}\label{sec:ablation}

This section presents a set of ablation experiments executed in the baseline architecture of the NAViDAd metric. The main objective is to fully understand the metric architecture by altering its main modules and observing the impact on the audio-visual quality prediction. The experiments were organized into four categories: 1) internal layers, 2) neuron distribution, 3) features selection, and 4) mapping strategy. The new models were trained using the same audio-visual content on which the NAViDAd metric was first trained.

\subsection{Internal Layers}

The first set of experiments examines the structure of the autoencoder module, more specifically the number of internal layers included in it. These experiments will determine how deep should the autoencoder be, so it keeps producing strong descriptive AV features. The \textit{Layers-0} configuration removes completely the autoencoder module from the baseline architecture. Basically, it takes the entire 115-by-$m$ matrix (containing the audio and video descriptive features) and passes it directly to the classification module to map the quality scores. The \textit{Layers-3} configuration adds one extra layer to the baseline architecture (115-60-25), containing a neuron distribution of the form 115-90-40-25. For this configuration, the basic input and output dimensions remain the same. That is, the model receives a 115-by-$m$ matrix and produces a 25-by-$m$ matrix. Finally, the \textit{Layers-4} configuration adds two extra layers to the baseline architecture (115-60-25), containing a neuron distribution of the form 115-100-70-50-25. Again, both input and output dimensions were kept the same.

\subsection{Neuron Distribution}

The second set of experiments analyzes the autoencoder structure,  focusing on the dimensions of the AV  features produced by the autoencoder module. Our goal is to determine how the dimensions of the AV features produced by the autoencoder affect the metric performance. In other words, how many AV features should be passed to the classification module to produce accurate audio-visual quality predictions. For this analysis, two setups of neuron’s distributions were considered.

The \textit{Nodes-10} configuration modifies the basic neuron distribution of the autoencoder module to produce a 10-by-$m$ matrix. This new distribution (115-60-10) reduces the number of AV features produced by the autoencoder from 25 to 10. With this reduction, we want to test if fewer features can reduce the model performance. The \textit{Nodes-50} configuration, on the other hand, increases the number of AV features and produces a 50-by-$m$ matrix. Its neuron distribution (115-60-50) duplicates the original number of AV features produced by the autoencoder module. The goal is to test if more features improve the model's performance.

\subsection{Features Selection}
\vspace{-0.1cm}
This set of experiments analyzes the effect of using only visual or audio features to train the model.  The \textit{Visual Features} configuration uses only the visual descriptive features for training. It takes a 90-by-$m$ matrix, containing the NSS features and the spatial and temporal features, to train the model. For this experiment the neuron distribution is 90-50-20, i.e., a 20-by-$m$ matrix is produced by the autoencoder and, then, it is passed to the classification module.  The \textit{Audio Features} configuration uses only the audio descriptive features to train the model. More specifically, the model is trained using the 25-by-$m$ matrix, which is the spectrogram representation of the audio signal. The model uses a neuron  distribution of the form 25-18-10, which means the autoencoder module produces a 10-by-$m$ matrix to feed the classification module. 

\subsection{Mapping Strategy}

The last experiment focuses on the classification module of the NAViDAd metric. For this experiment, the softmax function is substituted by a support vector regression (SVR) module. The SVR configuration maintains the same autoencoder architecture used in the baseline model (115-60-25). The SVR module is trained using the 25-by-$m$ matrix produced by the autoencoder module and the target set, which contains the quality scores associated with the audio-visual content. Once trained, the SVR module is able to map the AV features into a quality score in the range of $<$1,5$>$. Our goal is to determine if the softmax approach is a best alternative for mapping the AV features into quality scores. It also explores the impact of this module on the overall performance of the NAViDAd metric.

\section{Simulation Results and Discussion}\label{sec:discussion}


Table \ref{tab:results} presents the Pearson and Spearman correlation coefficients (PCC and SCC), along with the root mean square errors (RMSE), gathered from testing the different ablation experiment configurations on the audio-visual database. The results were organized according to the configuration categories presented in Section \ref{sec:ablation} and grouped by audio and video types of distortion and the overall scores. Additionally, Figure \ref{fig:ablation} presents a set of boxplots depicting the distribution of the results across all test iterations. Next, we discuss the results for each one of the experiments.

\subsection{On the impact of the number of internal layers}

As mentioned earlier, three configurations were selected to train different models with the goal of determining the adequate number of internal neuron layers required  to produce accurate quality predictions. We start by analyzing the performance of the \textit{Layers-0} configuration, i.e., a model trained without the autoencoder module. From Table \ref{tab:results}, it can be observed that the model performance drops when compared to the baseline configuration (from 0.88 to 0.86 for the PCC). However, this is not a significant loss in terms of accuracy, and it suggests that a softmax function is capable of predicting the audio-visual quality using raw audio and video features. Nevertheless, a light model can be  useful for implementations when less computational power is available.

As for Layer-3 configuration, an improvement in the performance is observed, when compared to the baseline model (0.88 to 0.91 for the PCC). These results indicate that, by adding one more layer to the original autoencoder, stronger AV features are obtained, leading to more accurate quality predictions. This type of configuration might be useful in a context where more accurate responses are needed and extra computational power (during the training) is available.
Layers-4 configuration presented a similar performance to the Layers-3. For this configuration, two extra layers were included to the baseline autoencoder module. By observing the results of this configuration, it can be inferred that just one extra layer is enough to improve the baseline performance without unnecessary training effort. 

\subsection{On the impact of the internal neuron distribution}

As mentioned earlier, this experiment had the objective of finding a suitable dimension for the AV descriptive features produced by the autoencoder module.  The Nodes-10 configuration was trained to produce a 10-by-m matrix as AV descriptive features. From Table \ref{tab:results}, it can be observed that there were no major changes in the overall performance of the model when compared to the baseline results. Based on these results, it can be assumed that an output set of 10 features (produced by the autoencoder module) retains an equivalent amount of information that the 25 set of features produced by the baseline design of the module. This property can be very useful for the design of a lighter audio-visual metric.

The second configuration was the Nodes-50. This model was trained to produce a 50-by-m matrix representing the AV descriptive features. The model showed no difference in terms of performance when compared to the Nodes-10 and baseline configurations. This means that a larger set of AV features (passed to the classification module) do not have a major impact in the performance of the module.

\subsection{On the impact of the features selection}

Two models were trained using either just visual features or audio features to find out how important are these features for the overall performance of the metric. Surprisingly, the Visual Features (VF) model presented a slightly better performance when compared to the baseline model design (0.88 to 0.90 for PCC). These results suggest that removing the audio features from the training input does not affect the model’s performance. In fact, it produces more accurate predictions in terms of audio-visual quality. Such behaviour might indicate that visual degradations dominated the overall audio-visual perception of quality.

When tested, the Audio Features (AF) model presented a considerable drop in terms of performance (0.52 for PCC). In accordance with the results of the VF model, removing the visual features from the training affected severely the model’s performance. 

\subsection{On the impact of the mapping strategy}

This last experiment investigates the impact that the classification module has in the overall performance of the baseline metric design. To do so, a model is trained replacing the softmax function by an SVR module while maintaining the autoencoder module original design. Results showed very poor performance of the SVR model (0.30 for PCC). Such performance evidence how important is the softmax function in the design of the baseline model. Even with a good set of AV features an effective mapping strategy is key to achieve accurate audio-visual quality predictions.

{In addition to the valuable insights presented in these experiments regarding the NAViDAd architecture, additional generalisable conjectures can be drawn from these results for the development of audio-visual quality metrics in general. It was observed from the results on the impact of feature selection that video features were capable of predicting the overall quality of content with audio and video distortions. These results are similar to those reported in \cite{martinez2019no}, where a NR video metric (which shares a common architecture with NAViDAd) was tested on an audio-visual database containing only video distortions (Packet loss and Frame-Freezing). This might imply that, in the presence of certain types of visual distortions, some audio distortions exert no influence in the overall perceived quality and only video features are required to obtain accurate quality predictions. In line with these results, a model using only the audio features was not capable of predicting the perceived quality in the presence of audio and video degradations. However, when the NR audio metric presented in \cite{phdthesisHelard2019} (also sharing NAViDAd’s architecture) was tested on an audio-visual database containing just audio distortions, the performance was better (0.75 for PCC). This lead us to infer that in the presence of only audio degradations (i.e. audio accompanied by distortion free video)  audio features can predict the overall audio-visual quality.}

\section{Conclusions}\label{sec:conclusions}

In this paper, a set of ablation experiments were carried out using the base architecture of the NAViDAd metric. This metric is a blind audio-visual quality model based on a deep autoencoder technique. The ablation experiments were organized in four categories and they studied different aspects of the NAViDAd architecture.    The first experiment examined the performance of different models by altering the number of internal layers in the autoencoder.
It was observed that three layers were sufficient to provide accurate predictions without increasing the complexity of the metric. The second experiment explored how different AV features dimensions affected the overall performance of the metric. It was observed that only 10 AV features (instead of 25 in the original design) were enough to obtain an equivalent performance to the baseline model. The third set of experiments revised the performance of the model when only audio or visual features were used for training. It was observed that just the visual features were able to produce accurate audio-visual quality predictions. Finally, the last experiment explored the impact that replacing the classification module by an SVR model. Results showed that the classification module is key in the metric design. {Further analysis can be made in order to optimize the training parameters that were not considered for this study (Table \ref{tab:params}).}

Overall, these sets of ablation experiments provided with valuable feedback that allow us to fully understand the metric architecture. Moreover, the information gathered in this study will permit us to design a more sophisticated version of the NAViDAd metric. {The  results from this study validate the architecture components of NAViDAd. It was observed that leaving out the autoencoder module has less of an affect on the performance than substituting the classification module. Changing the classification module had a significant impact on the overall model performance.} Ablation studies face a challenge with limitless potential test configurations. This chosen experiments explored the number of internal model layers, neuron distributions, feature selection and mapping strategies and the results present a rather wide picture of the problem, bringing up different aspects that need to be considered for designing better quality prediction models. 



\bibliographystyle{IEEEtran}
\bibliography{bibliografia}


\end{document}